\documentclass[prb,twocolumn,showpacs,preprintnumbers,amsmath,amssymb]{revtex4}
\usepackage{dcolumn}
\usepackage{bm,amsmath,amssymb}
\usepackage{graphicx}
\usepackage{bm}
\begin{document}
\pacs{75.30.Kz, 64.60.My, 64.70.P-, 75.47.Lx}
\title{Conversion of glassy antiferromagnetic-insulating phase to equilibrium ferromagnetic-metallic phase by devitrification and recrystallization in Al substituted Pr${_{0.5}}$Ca$_{0.5}$MnO${_3}$}

\author{A. Banerjee, Kranti Kumar and P. Chaddah}
\affiliation{UGC-DAE Consortium for Scientific Research\\University Campus, Khandwa Road\\
Indore-452017, M.P, India.}
\date{\today}
\begin{abstract}
We show that Pr${_{0.5}}$Ca$_{0.5}$MnO${_3}$ with 2.5\% Al substitution and La${_{0.5}}$Ca$_{0.5}$MnO${_3}$ (LCMO) exhibit qualitatively similar and visibly anomalous M-H curves at low temperature. Magnetic field causes a broad first-order but irreversible antiferromagnetic (AF)-insulating (I) to ferromagnetic (FM)-metallic (M) transition in both and gives rise to soft FM state. However, the low temperature equilibrium state of Pr$_{0.5}$Ca$_{0.5}$Mn$_{0.975}$Al$_{0.025}$O$_3$ (PCMAO) is FM-M whereas that of LCMO is AF-I. In both the systems the respective equilibrium phase coexists with the other phase with contrasting order, which is not in equilibrium, and the cooling field can tune the fractions of the coexisting phases. It is shown earlier that the coexisting FM-M phase behaves like `magnetic glass' in LCMO. Here we show from specially designed measurement protocols that the AF-I phase of PCMAO has all the characteristics of magnetic glassy states. It devitrifies on heating and also recrystallizes to equilibrium FM-M phase after annealing. This glass-like AF-I phase also shows similar intriguing feature observed in FM-M magnetic glassy state of LCMO that when the starting coexisting fraction of glass is larger, successive annealing results in larger fraction of equilibrium phase. This similarity between two manganite systems with contrasting magnetic orders of respective glassy and equilibrium phases points toward a possible universality.
\end{abstract}
\maketitle
\section {Introduction}
Complexity of strongly correlated electronic systems has many intriguing manifestations in transition metal oxides \cite {dag1} and pervoskite manganites are one amongst them \cite{tokura1}. The prototype charge ordered (CO) half doped manganite, Pr${_{0.5}}$Ca$_{0.5}$MnO${_3}$, undergoes a transition to CO insulating (I) state with antiferromagnetic (AF) order on lowering temperature. This charge order melts into a charge liquid metallic (M) state, with ferromagnetic (FM) order, under application of large magnetic field. It has been shown that minimal quench disorder in the magnetic lattice of Pr${_{0.5}}$Ca$_{0.5}$MnO${_3}$ in the form of 2.5\% substitution of Al in Mn site \cite{sunil2} weakens the so called robust CO and `finite size clusters' are formed \cite{sunil1}. It is shown that AF-I state of this system, Pr$_{0.5}$Ca$_{0.5}$Mn$_{0.975}$Al$_{0.025}$O$_3$ (PCMAO), can be converted to a FM-M state by moderate fields at low temperature \cite{ban1}. However, the system does not get back to its original zero-field cooled (ZFC) state (AF-I) when the field is withdrawn isothermally. This naturally brings up the question about the equilibrium phase; is the ZFC state (AF-I) the equilibrium phase or is the remnant state (FM-M) the equilibrium phase? Similar irreversibility in isothermal magnetization is observed in other half-doped manganites where it is proved that equilibrium phase at low temperature is AF-I \cite{ban2, rawat1}. Significantly, it is revealed from specially designed high field measurement protocols that this minimal substitutional disorder radically changes the low temperature AF-I equilibrium phase in the parent compound to FM-M in PCMAO \cite{ban1, ban2}. `Cooling and heating in equal field' (CHEF) shows that a fraction of AF-I phase transforms to FM-M phase with the decrease in temperature and the low temperature state shows coexisting FM-M and AF-I phases. However, a specially designed `cooling and heating in unequal fields' (CHUF) protocol is used to clearly remove the ambiguity about the low temperature equilibrium phase. 

While the AF-I and FM-M phases in half-doped manganites are separated by a first order transition in the space of field (H) and temperature (T) control variables, the low temperature equilibrium phase is often masked by the glass-like arrest of the kinetics of the phase transformation. The arrested high temperature AF-I phase persists untransformed when PCMAO is cooled in zero field to low temperatures. Cooling in higher fields shows the AF-I to FM-M transformation with lowering temperature; the transition is hysteretic during warming. The fraction of the arrested AF-I phase can be tuned by cooling the sample in different fields (but at the same cooling rate), and this coexists with transformed FM-M phase at low temperature \cite{ban1, ban2}. 

We present in this paper detailed measurements on PCMAO to show that the glass like arrested AF-I state is similar to the `magnetic glass' state recently reported in La${_{0.5}}$Ca$_{0.5}$MnO${_3}$ (LCMO) \cite{pc1}, and can also be similarly devitrified and re-crystallized. The interesting contrast is that in LCMO the arrested state has FM order and is metallic, whereas in PCMAO the arrested state has AF order and is insulating. In section II, we shall give details of our sample and the measurement methodology. Results and discussions, section III, starts with depicting qualitatively similar but clearly anomalous isothermal field cycling effect on the magnetization (M) of both PCMAO and LCMO. In section IIIA, we shall report magnetization in varying temperature, in various fixed fields. We show that an AF-I to FM-M transformation, with hysteresis characteristic of a first order transition, is observed for H $\geq$ 1 Tesla. Simultaneously, hysteresis characteristics of a first order transition will be shown in resistivity measurements. In these measurements we follow the CHEF protocol. In section IIIB, we present magnetization measurements when we follow the CHUF protocol. A special case of this protocol is the commonly used zero field-cooled (ZFC) protocol; our general version allows the heating field H$_h$ to be smaller or larger than the cooling field H$_c$. Our measurements of magnetization as a function of temperatue show topologically different features depending on the sign of (H$_h$-H$_c$). We shall link this with our proposed phenomenology \cite{pc2, kranti}. We have recently reported that while the arrested glass-like FM-M state in LCMO devitrifies on heating, cooling after annealing causes better recrystallization to AF-I state \cite{pc1}. In section IIIC, we shall show similar results in PCMAO. The contrast with LCMO is that, here the arrested state is AF-I and recrystallized state is FM-M. We shall then show that, like in LCMO \cite{ban3}, successive annealing causes better recrystallization. Finally, we show in section IIID, that if the starting glass-like AF-I fraction is larger, then similar successive annealing produces a larger recrystallized equilibrium FM-M fraction. This apparently counterintuitive result has been reported earlier in LCMO \cite{ban3}, and we shall comment on possible implication of this general observation.

\maketitle
\section{Experimental details}
In this study we have used the same Pr$_{0.5}$Ca$_{0.5}$Mn$_{0.975}$Al$_{0.025}$O$_3$ sample as in Ref. \cite{sunil2, sunil1, ban1, ban2}. Preparation and characterization details of this sample can be found in Ref. \cite{sunil2}. The La${_{0.5}}$Ca$_{0.5}$MnO${_3}$ sample is the same as used in Ref. \cite{pc1, ban3}. The magnetization and resistivity measurements are performed using commercial setups (14 T physical property measurement system- vibrating sample magnetometer (PPMS-VSM), M/s. Quantum Design, USA). All the temperature variations are done at the fixed rate of 1.5 K/min. This is relevant because the cooling rate dependence of glass-like arrest phenomena will not be invoked here.

\maketitle
\section{Results and discussions}
Figures 1(a) and 1(b) show the effect of magnetic field cycling on the magnetization of the PCMAO and LCMO samples respectively, at 5 K after cooling in zero field. Both the samples show field induced broad first-order transition from AF to FM state in moderate fields ($>$ 5.5 T for PCMAO and $>$ 8.5 T for LCMO) in the initial field increasing cycle. As the field is increased towards the maximum available field of 14 T, the magnetization approaches the spin-aligned value of 3.5 $\mu$$_B$/Mn expected for these manganites around half doping. However, neither the field decreasing cycle shows the reverse first-order FM to AF transition, nor the next field increasing cycle shows the AF to FM transition observed in the original state. It is rather interesting that the subsequent field cycling portray M-H resembling soft FM state for both the samples. It may be noted that the initial increase in magnetization of LCMO, immediately above zero field is because of the untransformed FM phase which got arrested while cooling in zero field \cite{pc1}. Thus both the samples show qualitatively similar anomaly in isothermal magnetization.

\subsection{Cooling and heating in equal field (CHEF)}
To probe the above mentioned anomalous state we measure the magnetization of PCMAO, while cooling and heating in equal field. Figure 2(a) shows the magnetization in different fields following the CHEF protocol. This clearly shows broad first-order AF to FM transition with decrease in temperature. This is in clear contrast to the observations made in LCMO, where FM to AF transition is observed during cooling \cite{pc1, loudon}. However, even in 8 T field, fully saturated moment value is not achieved much below the closure of the thermal hysteresis. This indicates that the AF phase is not fully transformed to FM phase at the lowest temperature. This can be clearly observed in the inset of Fig. 2(a) where, along with the M-H curves of 5 K, the magnetization values at 5K in different fields after cooling in the same fields are plotted. All the field-cooled magnetizations in cooling and measuring fields between 1 to 8 T shows moment values below the return envelope curve, indicating existence of different fractions of the untransformed AF phase. This fraction decreases in higher cooling fields and fully transformed FM phase is obtained when the system is cooled in H $\geq$ 10 T. It is interesting to note that cooling in fields lower than the field required to cause the first-order field induced transition at 5 K ($\approx$ 5.5 T) also convert a fraction of the AF phase resulting in moment values higher than that of the initial field-increasing state.

Figure 2(b) show the resistivity of PCMAO in different fields measured following the CHEF protocol. These show broad first-order insulator to metal transition with decrease in temperature. However, anomaly similar to magnetization is also observed in resistivity in isothermal field variation at 5 K, which was shown recently in Ref. \cite{ban1}. Hence, similar to LCMO, coexisting phases with different fractions of AF-I and FM-M phases can be created at low temperature in PCMAO \cite{ban1, ban2}, one may be in equilibrium and the other in the `magnetic glassy' state. The above mentioned measurements following the CHEF protocol shows that the AF-I phase partially transforms to FM-M on cooling, which is the converse of what is observed in LCMO. However, CHEF does not bring out the glass-like arrested character of the coexisting AF-I phase fraction.

\subsection{Cooling and heating in unequal fields (CHUF)}
In this section we show that the nature of coexisting AF-I and FM-M phases at low temperature in PCMAO can be identified by the uncommon measurement protocol, CHUF. As mentioned earlier, heating in finite fields of the zero-field cooled (ZFC) state is a special case of CHUF and data is shown in Fig. 3. After cooling in zero-field, different field is applied each time while measuring the magnetization during heating. All these curves show initial increase with the increase in temperature, and the increase is sharper for higher fields. CHEF protocols have not shown any increase at these low temperatures for the corresponding measurement fields (Fig. 2(a)). Moreover, the magnetization value at 5 K is much less compared to the value achieved when cooled in the same measurement fields (Fig. 2(a)). It is shown in Fig. 2(a) that cooling and measuring in field $<$ 10 T renders a fraction of AF-I phase untransformed at 5K. Figure 3 shows much larger fraction of untransformed AF-I phase when the magnetization is measured in the corresponding fields after ZFC, consistent with the inset of Fig. 2(a). The sharp increase in magnetization of this state with the increase in temperature clearly indicates the rapid transformation of the untransformed AF-I phase fraction to FM-M phase. Such rapid transformation to the low temperature equilibrium phase when thermal energy is imparted to the system is akin to devitrification, which is an evidence of glassy state \cite{greer}. Though this state shows transformation to the FM-M state with time (not shown here), a more visual and informative (as discussed below) manifestation of devitrification on the magnetization is shown following the CHUF protocol in Figs. 4 and 5.

In Fig. 4 we show CHUF measurements of magnetization, comparing data for the same value of H$_c$ but different values of H$_h$. We have data for various values of H$_c$, but show only representative data for (a) H$_c$ = 3 T and (b) H$_c$ = 2.5 T. For all H$_c$, de-arrest of AF-I to FM-M is observed only when H$_h$ $> $H$_c$. Similar dependence of CHUF magnetization on the sign of (H$_h$-H$_c$) was used in the Ref. \cite{ban2} to distinguish the equilibrium phase from the glass-like arrested coexisting phase in different manganites around half doping, highlighting the novelty of CHUF. The sharp rise in magnetization at low temperature resulting from de-arrest of the AF-I phase is also observed for ZFC measurements. The arrested AFI fraction is fixed by H$_c$, but as shown by Fig. 4(b) the extent of de-arrest (or rise in magnetization) is more when H$_h$ is more. This feature is observed for all values of H$_c$, including the data for H$_c$ = 0 in Fig. 3. This feature is consistent with T$_K$, the temperature for glass-like arrest, falling as H rises \cite{ban1, ban2}. 

It was shown earlier for variety of systems that broad H and T induced first-order transitions can be represented by quasi-continuum of supercooling (T*) lines, corresponding to spatial distribution of regions in the sample of the dimension of the correlation length, forming into T* bands in the H-T space. Consequently, H-T dependent glass-like arrest of kinetics was also represented by T$_K$ band in the H-T space. The CHUF magnetization is related to sign of the slopes of T* and T$_K$ bands in the H-T space for this case. Significantly, it was proposed that the nature of correlation between the T* and T$_K$ bands can be revealed from the CHUF measurements \cite{pc2}. We now discuss such measurements for PCMAO.

In Fig. 5 we show two representative examples of CHUF with (a) H$_h$ = 3 T and (b) H$_h$ = 4 T. The features outlined below have also been observed for various other values of H$_h$. We again find that de-arrest is observed whenever H$_c$ $<$ H$_h$, but not when H$_c$ $>$ H$_h$, corresponding to an equilibrium FM-M  phase at low temperature. We note that as  (H$_h$ - H$_c$) becomes smaller in magnitude, de-arrest is initiated at higher temperature. This was also observed in La-Pr-Ca-Mn-O \cite{kranti} and corresponds to the T* and T$_K$ bands being anti-correlated; i.e. regions with above-average T* have below-average T$_K$. It is rather intriguing, that such anti-correlation was shown to be universal for all other systems where the `magnetic glassy' states are identified \cite{kranti, roy1, rawat2}. Evidence of the anti-correlation is clear in Fig. 5, where the CHUF measurements show that for fixed H$_h$, de-arrest is initiated at higher temperature for higher H$_c$. This is exactly as predicted by the phenomenology for an FM equilibrium and AF arrested phase (see Fig. 3(c) of Ref. \cite{kranti}). We stress that the CHUF measurements not only allows to identify both the existence and magnetic order of the  `magnetic glassy' state but goes deeper to probe correlation between T* and T$_K$ without taking recourse to time domain \cite{wu}.                 

\subsection{Recrystallization of glass-like AF-I phase by single and successive annealing}
Unambiguous and rather visual evidence of the coexisting AF-I `magnetic glassy' state in PCMAO is given in the previous section from devitrification. In this section, we show that a fraction of this magnetic glass can be recrystallized by single annealing and a larger fraction by successive annealing to FM-M phase analogous to the conventional structural glass \cite{greer}. Figure 6 shows this recrystallization of the AF-I phase fraction to FM-M by single annealing through magnetization measurement. Cooling the sample from room temperature in 2 T results in larger fraction of AF-I phase at 5 K, a fraction of which gets devitrified to FM-M phase indicated by sharp increase in magnetization measured while warming after isothermally changing the field to 5 T at 5 K. However, instead of heating all the way up to room temperature, the sample is heated only up to 60 K and cooled back again without changing the field. This cooling from 60 K results in a higher magnetization value at 5 K, more than the value found after cooling the sample from room temperature in 5 T (FCC). Thus a fraction of residual AF-I phase is recrystallized to FM-M phase upon annealing to 60 K producing larger magnetization in the same field (5 T) and temperature (5 K) than the corresponding FCC value. Similar recrystallization by single annealing can be observed in the inset of Fig. 6 when the initial cooling from room temperature as well as all the subsequent temperature cycling are performed in a fixed field of 5 T. It is obvious from this figure that there is significant increase in magnetization after annealing to 60 K compared to its value after the initial cooling.

Now we show more effective recrystallization of AF-I magnetic glassy fraction by successive annealing through both resistivity and magnetization measurements. Figure 7 shows large resistivity of PCMAO at 5 K when cooled and measured in 2.1 T field. Thereafter the resistivity is measured in the same field of 2.1 T while heating and cooling, each time annealing at progressively higher temperatures. Resistivity of only a few such temperature cycling is shown for clarity. It is evident that the resistivity at 5 K decreases after each annealing at successive higher temperatures up to the annealing temperature of 60 K indicating recrystallization of arrested AF-I phase to FM-M phase. However, the increase in resistivity after annealing at 80 K occurs because of the reverse transformation of FM-M phase fraction to AF-I phase on approaching superheating spinodal of first-order transformation. This trend continues for other temperatures $>$ 80 K (not shown here). Qualitatively similar effect of recrystallization by successive annealing are observed in resistivity measurements in many other fixed cooling and annealing fields. This is the converse of our report on LCMO where the glassy arrested phase is FM-M and the equilibrium phase is AF-I. 

Figure 8 shows recrystallization by successive annealing through measurement of magnetization while cooling and annealing in fixed field of 5 T, each time traversing to progressively higher temperatures. It is shown in Fig. 2(a) that, a fraction of the high temperature AF-I phase remains untransformed at 5 K when cooled in 5 T. Now, if this glass-like arrested state is subjected to successive annealing then recrystallization to equilibrium FM-M phase takes place showing concomitant increase in magnetization at 5 K. It needs to be reiterated that such recrystallization in LCMO occurs from glass like arrested FM-M to equilibrium AF-I phase, whereas in the present case of PCMAO the phases appear in reverse order.

\subsection{More glass gives better crystal}
In this section we give evidence of an intriguing but unequivocal experimental observation that is rather counterintuitive and significant. We show that when the starting fraction of the glassy phase is higher, successive annealing in same field produces larger fraction of equilibrium (crystal) phase at low temperature. This is demonstrated in Fig. 9 by measuring magnetization at fixed temperature (5 K) after each annealing in the fixed field (5 T) at the successively higher temperatures. Two initial states with different fractions of arrested AF-I phases are produced at 5 K by cooling in 2 and 5 T. Then after isothermally changing the field to 5 T for the 2 T cooled state, magnetization is measured at 5 K for both these states after successive annealing at progressively higher temperatures without changing the field. In Fig. 9 magnetization at 5 K in 5 T is plotted as a function of annealing temperatures. As shown in Fig. 2 (a), the 2 T cooled state has larger fraction of AF-I phase at 5 K compared to the 5 T cooled state. When the field is isothermally changed to 5 T, this state gives much lower magnetization value at 5 K (see inset of Fig. 9) compared to the 5 T cooled state. The magnetization value of this state at 5 K remains lower than the 5 T cooled state even after successively annealing up to 20 K (see inset of Fig. 9). After next annealing to 30 K, the magnetization of both the states become almost equal. Significantly, annealing to subsequent higher temperatures ( up to 70 K) show drastic effect, 2 T cooled state has higher magnetization than the 5 T cooled state in same measurement field and temperature. Thus, when the starting fraction of glass-like arrested AF-I phase is higher, successive annealing produces larger fraction of crystal-like equilibrium FM-M phase in the same measurement condition. Similar intriguing effect has been observed in LCMO for different measurement fields as well as for different initial fraction of glass \cite {ban3, ban4}, though in that case the glassy phase is FM-M and here the glassy phase is AF-I.                 
\maketitle
\section{Conclusions}
We conclude that all the characteristic features of the FM-M magnetic glassy state of LCMO also exist for the kinetically arrested AF-I phase in PCMAO. We bring out the similarities as well as contrasts between the two manganites around half-doping giving below the highlights of the present study:
(i)    PCMAO and LCMO show qualitatively similar but anomalous isothermal M-H behaviors at low temperature. Both show field induced broad first-order transition from AF-I to FM-M state which is not reversible. The resulting FM-M state is a soft ferromagnet without observable hysteresis. 
(ii)  PCMAO shows temperature induced first-order transition from AF-I to FM-M state with decrease in temperature but opposite transition from FM-M to AF-I state is observed for LCMO, indicating contrasting nature of  the equilibrium phase. However, in both the systems the complimentary order coexists with the equilibrium phase fraction at the lowest temperature.
(iii) A specially designed measurement protocol, CHUF, can unravel the nature of the coexisting non-equilibrium phase, which is found to be glass-like. Significantly, CHUF allows probing deeper into the phenomenon of `magnetic glass' and brings out an anti-correlation between superheating and glass transition temperatures of different regions of the sample.
(iv) This glass-like AF-I state in PCMAO resembles the conventional glass. Moreover, similar to magnetic glass of LCMO (having opposite nature of magnetic order), it devitrifies on heating, recrystallizes after annealing and more effectively by successive annealing. 
(v) Most interesting as well as intriguing observation of this study is that larger fraction of equilibrium phase can be recrystallized when the starting equilibrium fraction is smaller in the coexisting phase. In other words, more glass makes better crystal both in PCMAO and LCMO. 

Thus, there appears to be a universality in the magnetic glassy state irrespective of its magnetic order or electron transport properties. This needs to be explored for other systems and with other experimental techniques. It also needs to be checked whether this conclusion of ``more glass gives more crystal'' is valid for structural glasses. We must record that traversing unconventional (H, T) paths is experimentally much easier than traversal of similar (P, T) paths that would be required for structural glasses. This is because H is varied at the sample without a medium, which does not complicate independent variation of the two control parameters. We thus expect systems like those studied here to shed more light on the physics of glasses.  

\maketitle\section{Acknowledgement}
DST Government of India is acknowledged for funding the 14 Tesla PPMS-VSM.

\begin{figure*}
	\centering
		\includegraphics{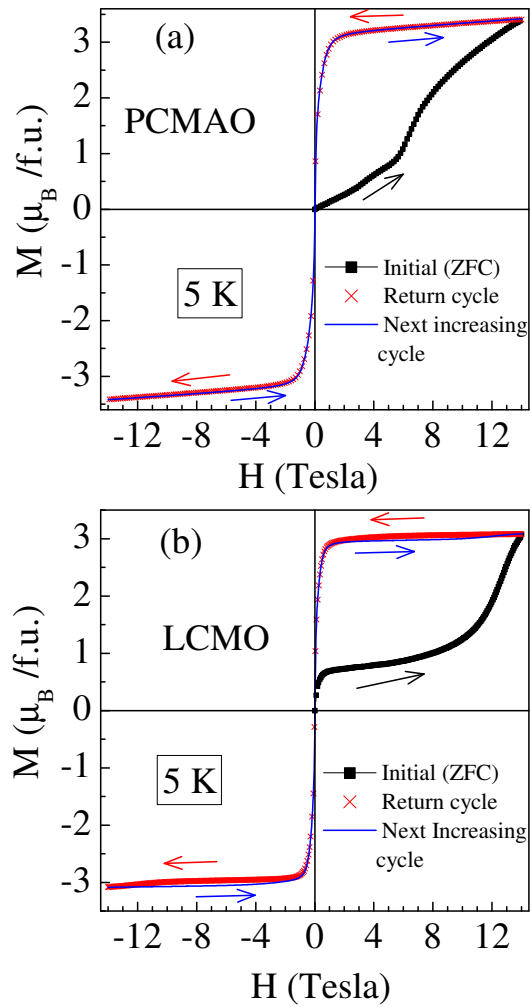}
	\caption{(Color online) M-H curves of 	PCMAO and LCMO at 5 K after cooling the samples from 320 K in zero field. (a) Initial field increasing cycle of PCMAO shows linear increase in magnetization in lower field for the AF state. Then it undergoes a broad field induced transition to FM state at higher fields and approaches the spin aligned moment value at 14 T.  However, the field decreasing cycle does not show the reverse transition to AF state indicating anomaly in the field induced first-order transition. With the decrease in H, magnetization decreases to zero resembling a soft FM state. This is evident in the subsequent field cycling which shows a hysteresis loop without any observable opening. (b) M-H of LCMO following same measurement protocol of (a). The initial increase in magnetization just above zero field is because of the presence of a fraction of the soft FM phase which got arrested during cooling in zero-field. Remaining AF phase fraction undergoes a field-induced transition to FM state; thereafter it shows soft magnetic behavior in subsequent field cycles akin to PCMAO. In both cases, the ZFC AF-I state cannot be retrieved after the field induced transition to FM-M state, unless the sample is warmed above the superheating temperature. }
	\label{fig:Fig1}
\end{figure*}

\begin{figure*}
	\centering
		\includegraphics{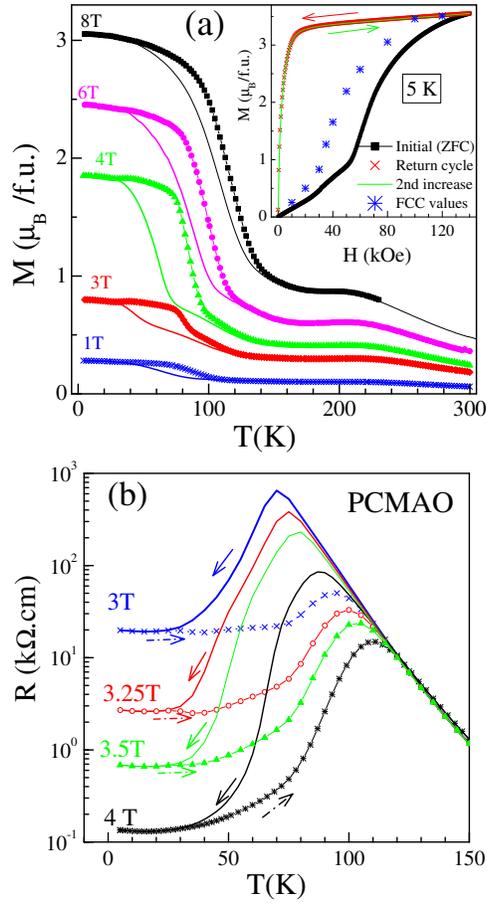}
	\caption{(Color online) Temperature induced broad first-order transition from AF-I to FM-M state of PCMAO in different fields. (a) The AF to FM transition does not appear to have transformed the complete sample to FM state even much below the closure of thermal hysteresis indicating incomplete first-order transformation process. The magnetization values at lowest temperature are much less than the values expected for such soft FM state. This is clearly brought out in the inset of (a) where the magntization values in different fields at 5 K after cooling in the same fields are plotted along with the M-H. For H $<$ 10 T these values fall between initial and return curves indicating coexistence of AF and FM phases in same field and temperature. (b) Shows the thermal hysteresis of resistivity in different fields from I to M state which are associated with the AF and FM phases respectively for PCMOA.}
	\label{fig:Fig2}
\end{figure*}

\begin{figure*}
	\centering
		\includegraphics{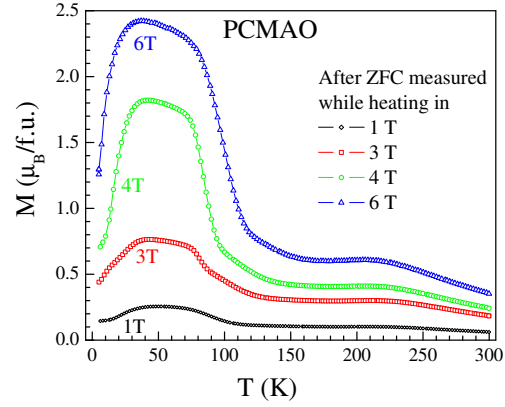}
	\caption{(Color online) Magnetization of PCMAO in different fields while warming after cooling in zero field. The sharp increase from lowest temperature signifies devitrification of the arrested AF-I phase to equilibrium FM-M phase. With further increase in temperature, magnetization shows a sharp fall arising from the reverse conversion of the FM-M to AF-I phase on approaching the superheating limit. Thus heating in field higher than the cooling field (zero in this case) gives rise to two sharp changes with opposite signs, one at low temperature because of devitrification to the low temperature equilibrium phase and another at higher temperature resulting from first-order transition to the high temperature equilibrium phase having contrasting order.}
	\label{fig:Fig3}
\end{figure*}

\begin{figure*}
	\centering
		\includegraphics{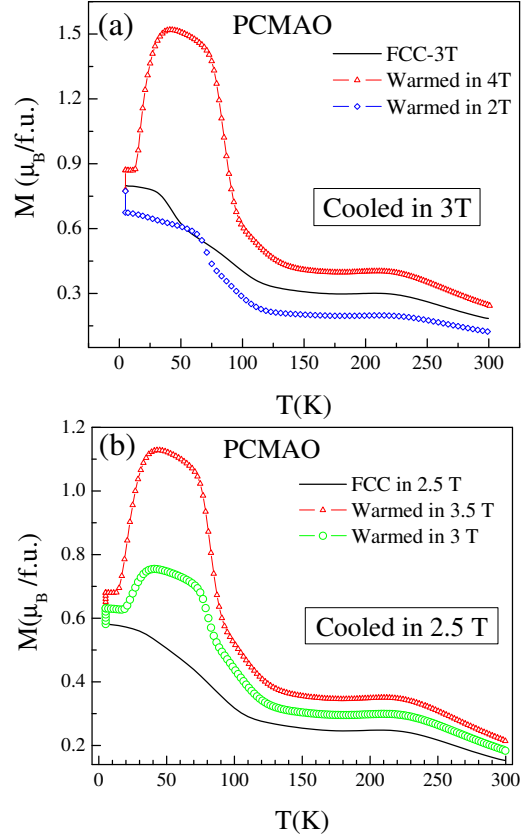}
	\caption{(Color online) Magnetization following CHUF protocol (defined in text) after cooling in fixed field but warming in unequal fields. (a) Shows magnetization in 2 and 4 T while warming after cooling in 3 T along with the 3 T cooling run. When the cooling field (H$_C$) is lower than the heating field (H$_h$), magnetization shows two sharp changes of opposite signs arising from devitrification followed by reverse conversion. Whereas when H$_C$ $<$ H$_h$ the change associated with devitrification is absent. (b) Shows magnetization in 3 and 3.5 T while warming after cooling in 2.5 T along with the 2.5 T cooling run. Here, both the warming fields (H$_h$) are higher than H$_C$ as result the magnetization in both cases shows two sharp but opposite changes. The sharp increase in magnetization associated with devitrification for the lower warming field (3 T) starts at higher temperature than the 3.5 T curve because the devitrification start on approaching the kinetic arrest band which, for this case (FM-M ground state), moves to higher temperature with decrease in field (see Refs. \cite{ban1, ban2}).}
	\label{fig:Fig4}
\end{figure*}

\begin{figure*}
	\centering
		\includegraphics{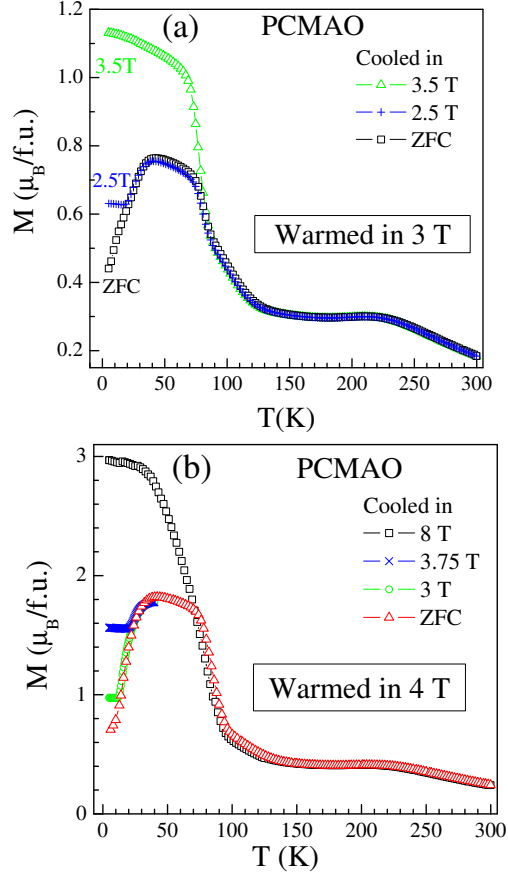}
	\caption{(Color online) Magnetization while warming following CHUF protocol where H$_C$ is varied but H$_h$ is kept fixed. (a) Cooling fields are 0, 2.5 and 3.5 T and the warming field (H$_h$) is 3 T. It is clear that when H$_C$ $<$ H$_h$ (in case of H$_C$ = 0 and 2.5 T) there are two sharp changes whereas for H$_C$ = 3.5 T  there is only one sharp change at higher temperature because of reverse transformation to AF-I state. (b) In this case the H$_h$ is fixed at 4 T after cooling in 0, 3, 3.75 and 8 T. The behavior is consistent with the sign of (H$_C$ - H$_h$). Devitrification starts at lower temperature for H$_C$ = 3 T than for H$_C$ = 3.75 T. }
	\label{fig:Fig5}
\end{figure*}

\begin{figure*}
	\centering
		\includegraphics{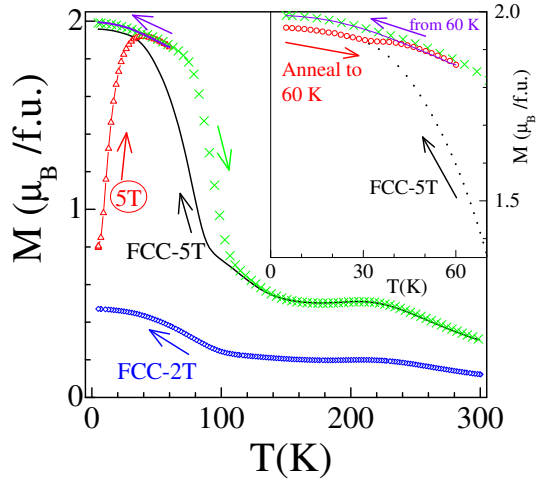}
	\caption{(Color online) Recrystallization of the FM-M phase in PCMAO after single annealing at 60 K is shown through magnetization measurements. When the sample is cooled in 2 T the magnetization at 5 K is small because of the presence of large fraction of arrested AF-I phase. After isothermally changing the field at 5 K to 5 T the magnetization is measured during warming. The sharp increase indicates devitrification of a fraction of the arrested AF-I phase to FM-M phase. After warming to 60 K the sample is cooled back again. This second cooling after annealing to 60 K results in higher magnetization at 5 K in 5 T compared to its value when the sample is directly cooled in 5 T from 320 K. In the inset we show this clearly by cooling and annealing in the same field. For the inset, magnetization is measured while cooling in 5 T, then while warming from 5 to 60 K and again while cooling to 5 K followed by warming all the way. It is visibly clear that 2nd cooling after annealing to 60 K produces larger magnetization at 5 K than its value after initial cooling. This confirms recrystallization of the FM-M phase fraction after single annealing at 60 K.}
	\label{fig:Fig6}
\end{figure*}

\begin{figure*}
	\centering
		\includegraphics{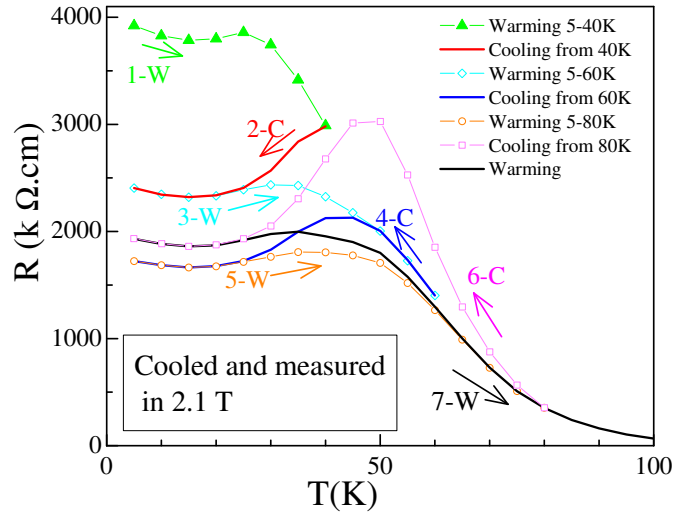}
	\caption{(Color online) Recrystallization of FM-M phase from glass-like arrested AF-I phase in PCMAO during successive annealing at progressively higher temperatures through resistivity measurement in 2.1 T. Resistivity shows a large value at 5 K after cooling in 2.1 T arising from larger fraction of AF-I phase. This value of resistivity at 5 K decreases substantially after annealing to 40 K. Further decrease in this value takes place after annealing to 60 K. Thereafter, this value shows an increasing trend when annealed to 80 K or higher because of reverse transformation to AF-I phase and also rules out any artifact related to irreversible change in the sample. Successive warming (W) and cooling (C) cycles are sequentially numbered along with the direction of the respective temperature excursion (denoted by W or C).}
	\label{fig:Fig7}
\end{figure*}

\begin{figure*}
	\centering
		\includegraphics{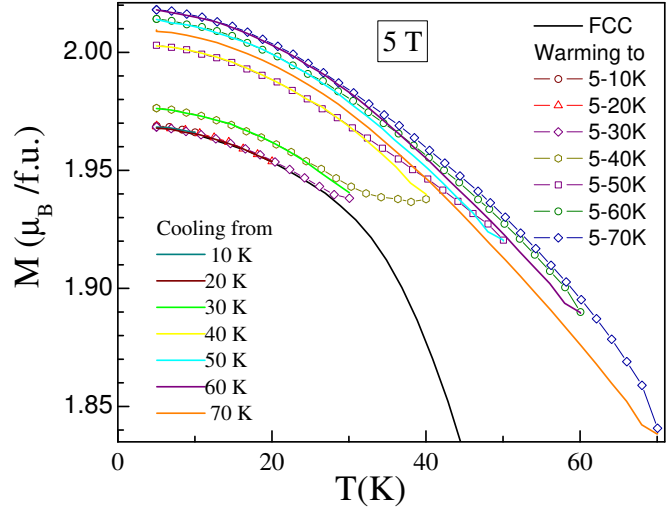}
	\caption{(Color online) Recrystallization of FM-M phase from glass-like arrested AF-I phase in PCMAO during successive annealing at progressively higher temperatures (increasing in the steps of 10 K) through magnetization measurement in 5 T. Magnetization is measured during cooling in 5 T, producing a smaller value at 5 K. Then cooling to 5 K after annealing at 10 and 20 K do not produce noticeable change. Thereafter, the magnetization at 5 K shows consecutively higher values with successive annealing up to 60 K. Then after annealing at 70 K or higher temperatures the magnetization at 5 K starts coming down because of reverse transformation.}
	\label{fig:Fig8}
\end{figure*}

\begin{figure*}
	\centering
		\includegraphics{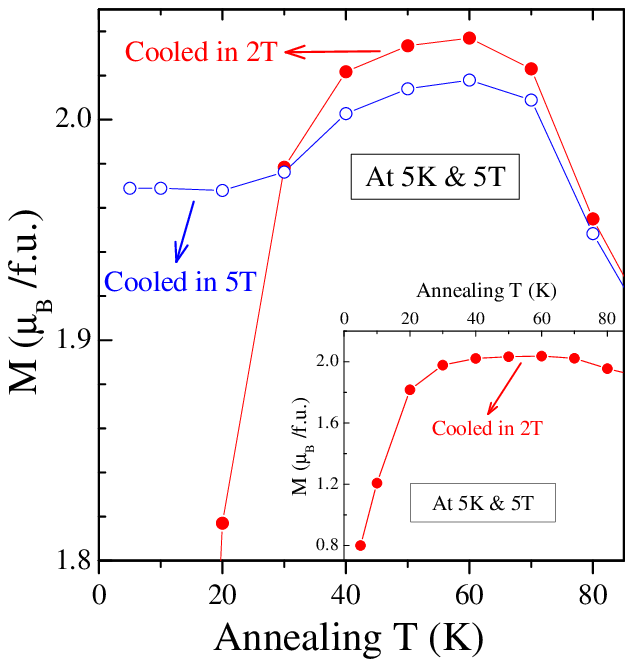}
	\caption{(Color online) Evidence of larger fraction of glass-like AF-I phase producing larger fraction of equilibrium (crystal-like) FM-M phase by successive annealing. Magnetization values measured in 5 T at 5 K after initial cooling in 5 T and after successively annealing at progressively higher temperatures in the same measurement field (5 T) are plotted as a function of respective annealing temperatures. The glass-like AF-I phase fraction in 5 T cooled state progressively recrystallize to FM-M phase during successive annealing up to 60 K. Thereafter, it start decreasing because of reverse conversion. The 2 T cooled state has larger fraction of arrested AF-I phase at 5 K resulting in much smaller magnetization (as shown in the inset) when the field is isothermally changed to 5 T. Even after annealing this states in the same measurement field (5 T) to 10 and 20 K the magnetization at 5 K in 5 T is much smaller than the corresponding values for the 5 T cooled state. However, after annealing to 30 K the magnetization at 5 K steeply rises to become equal and thereafter it shows a significant overshoot from the corresponding values of the 5 T cooled state with each annealing up to 60 K. This clearly shows that the presence of larger fraction of glass-like AF-I phase triggers conversion of larger fraction to FM-M phase.}
	\label{fig:Fig9}
\end{figure*}


\begin{thebibliography}{}
\bibitem[1]{dag1} Elbio Dagotto, Science \textbf{309}, 257 (2005); and references therein.
\bibitem[2]{tokura1} Y. Tokura, Rep. Prog. Phys. \textbf{69}, 797 (2006); and references therein. 
\bibitem[3]{sunil2}Sunil Nair and A. Banerjee, J. Phys.: Condens. Matter \textbf{16}, 8335 (2004).
\bibitem[4]{sunil1} Sunil Nair and A. Banerjee, Phys. Rev. Lett. \textbf {93}, 117204 (2004).
\bibitem[5]{ban1}A. Banerjee, K. Mukherjee, Kranti Kumar and P. Chaddah, Phys. Rev. B \textbf{74}, 224445 (2006).
\bibitem[6]{ban2}A. Banerjee, A. K. Pramanik, Kranti Kumar and P. Chaddah, J. Phys.: Condens. Matter \textbf{18}, L605 (2006).
\bibitem[7]{rawat1}R. Rawat, K. Mukherjee, Kranti Kumar, A. Banerjee and P. Chaddah, J. Phys.: Condens. Matter \textbf{19}, 256211 (2007).
\bibitem[8]{pc1} P. Chaddah, Kranti Kumar and A. Banerjee, Phys. Rev. B \textbf{67}, 100402(R) (2008).
\bibitem[9]{pc2}P. Chaddah, A. Banerjee and S. B. Roy, arXiv:0601095 (unpublished).
\bibitem[10]{kranti}Kranti Kumar, A. K. Pramanik, A. Banerjee, P. Chaddah, S. B. Roy, S. Park, C. L. Zhang and S. -W. Cheong, Phys. Rev. B {\bf73}, 184435 (2006).
\bibitem[11]{ban3}A. Banerjee, Kranti Kumar and P. Chaddah, arXiv:0711.2347 (unpublished); J. Phys.: Condens. Matter (In press).
\bibitem[12]{loudon}J. C. Loudon, N. D. Mathur and P. A. Midgley, Nature {\bf420}, 797 (2002).
\bibitem[13]{greer}A. L. Greer, Science {\bf267}, 1947 (1995).
\bibitem[14] {roy1}S. B. Roy, M. K. Chattopadhyay, A. Banerjee, P. Chaddah, J. D. Moore, G. K. Perkins, L. F. Cohen,  K. A. Gschneidner, Jr. and V. K. Pecharsky, Phys. Rev. B {\bf75}, 184410 (2007).
\bibitem[15]{rawat2}Pallavi Kushwaha, R. Rawat and P. Chaddah, J. Phys.: Condens. Matter {\bf 20}, 022204 (2008).
\bibitem[16]{wu}W. Wu, C. Israel, N. Hur, P. Soonyong, S. -W. Cheong and A. De Lozane, Nature Materials {\bf 5}, 881 (2006 ).
\bibitem[17]{ban4}A. Banerjee, Kranti Kumar and P. Chaddah, arXiv:0710.5585 (unpublished).


\end{thebibliography}
\end{document}